\documentclass[aps,10pt,showpacs,twocolumn]{revtex4}
\usepackage{amsmath, amssymb, lipsum}
\usepackage{graphics, graphicx, epsfig, color, subfigure}
\begin{document}

\title{Mean-field theory of baryonic matter for QCD in the large $N_{c}$ and heavy quark mass limits }

\date{\today}

\author{Prabal Adhikari}
\email{prabal@umd.edu}
\affiliation{Maryland Center for Fundamental Physics and the Department of Physics, 
University of Maryland, College Park, MD 20742-4111}

\author{Thomas D. Cohen}
\email{cohen@umd.edu}
\affiliation{Maryland Center for Fundamental Physics and the Department of Physics, 
University of Maryland, College Park, MD 20742-4111}

\begin{abstract}
We discuss theoretical issues pertaining to baryonic matter in the combined heavy-quark and large $N_c$ limits of QCD.  Witten's classic argument that baryons and interacting systems of baryons can be described in a mean-field approximation with each of the  quarks moving in an average potential due to the remaining quarks is heuristic.  It is important to justify this heuristic description for the case of baryonic matter since systems of interacting baryons are intrinsically more complicated than single baryons due to the possibility of hidden color states---states in which the subsystems making up the entire baryon crystal are not color-singlet nucleons but rather colorful states coupled together to make a color-singlet state.  In this work, we provide a formal justification of this heuristic prescription.  In order to do this, we start by taking the heavy quark limit, thus effectively reducing the problem to a many-body quantum mechanical system.  This problem can be formulated in terms of integrals over coherent states, which for this problem are simple Slater determinants.  We show that for the many-body problem, the support region for these integrals becomes narrow at large $N_c$, yielding an energy which is well-approximated by a single coherent state---that is a mean-field description. Corrections to the energy are of relative order $1/N_c$.  While hidden color states are present in the exact state of the heavy quark system, they only influence the interaction energy at sub-leading order in $1/N_{c}$.
\end{abstract}
\maketitle
\section{Introduction}

The 't Hooft large $N_{c}$ and the heavy quark mass limits provide a tractable regime to study the properties of baryons and baryonic matter~\cite{witten}. The analogous problem of nuclear matter in three-color QCD with realistic quark masses  cannot be solved analytically. Moreover, the system cannot be studied using conventional non-perturbative techniques such as lattice QCD, since at finite baryon density Monte Carlo methods break down due to the fermion sign problem~\cite{signproblem1,signproblem2}.  Thus, it is useful to consider tractable limits where some insights may be gained.  Here, we will consider baryons with heavy quarks in the large $N_c$ limit of QCD.  We note that there has been some recent interest in variants of the large $N_c$ limit with quarks in the two index anti-symmetric limit~\cite{AS1,AS2}.  In this work, we will use the conventional 't Hooft large $N_c$ limit with quarks in the fundamental representation.   We also assume that there are $N_f$ flavors of degenerate quarks.

A world with quarks much heavier than the QCD scale is comparatively simple.  The quarks are nonrelativistic and quark-antiquark pair production is suppressed.  The overall state must be a color-singlet.  Thus, for example, baryons contain $N_c$ quarks described by a many-particle wave function that is completely antisymmetric in space, spin, flavor and color.  Moreover, the interaction is described via a simple two-body color Coulomb force with spin-dependent forces suppressed.    Effects of multiple gluon exchanges are suppressed including all many-quark forces.   This simplified world still involves a complicated quantum many-body problem.  This problem simplifies considerably in the large $N_c$ limit.   As first discussed by Witten~\cite{witten}, one can describe the baryon via a mean-field description:  each quark moves in a mean-field potential created by the remaining $N_{c}-1$ quarks.

A quantitative study of large $N_{c}$ baryons and low density baryonic matter for QCD with heavy quarks in three spatial dimensions was conducted recently~\cite{3+1}.  The study was based on Witten's mean-field analysis.  The paper determined baryon masses of isolated baryons and interaction energies for low-density baryonic matter, which takes the form of a crystal composed of baryons that are distorted relative to the isolated baryons but has a color-singlet form.  The restriction to low density matter was technical in nature as the calculation greatly simplified if only nearest neighbor interactions between nucleons are relevant---as one expects to occur at low-enough densities. The fundamental approximation was the mean-field one justified by Witten.  However, it was noted in Ref. \cite{3+1} that Witten's arguments were quite heuristic and that theoretical work was needed to put in a more solid footing~\cite{3+1}.   One point raised in Ref. \cite{3+1} was that the mean-field ansatz is a very poor approximation to the exact many-body wave function for the color-Coulomb problem for both baryons and baryonic matter; the overlap between the mean-field and the exact wave functions does not approach unity as $N_c \rightarrow \infty$.   However, it was argued that despite this poor description of the wave function, the energy was approximated well.  The basic point was that if each quark had a component of order $1/N_c$, which was not in the mean-field state, the energy would be accurate to relative order $1/N_c$ while the wave function would have a correction of order $N_{c}^{0}$.  For the case of single baryons, a simple argument of how this occurs was developed.  It was based on the $N_c$ scaling of matrix elements of $n$-particle-$n$-hole states on top of the mean-field state.  The argument was fairly powerful, but involved an assumption on the ordering of two limits.  Moreover, the argument was not simple to generalize to the case of baryonic matter.  One reason for this is an intrinsic complication of baryonic matter as compared to single baryons: the possible contribution of ``hidden color  state'' contributions to the total wave function.  A hidden color state is one in which the wave function is not composed entirely of subsystems, which are color-singlet nucleons, but rather includes colorful states of baryon number unity coupled together to make an overall color-singlet~\cite{hidden1,hidden2,hidden3}.   The purpose of the present paper is to provide a theoretical justification for the mean-field approximation for baryonic matter---doing so automatically justifies the single baryon case.

There is strong reason to believe that the mean-field approximation (which by construction has no hidden color state components) accurately describes the interaction energy of baryon matter at leading order in $1/N_{c}$. One major reason for this is an extension of the analysis to low density baryonic matter for large $N_c$ QCD with heavy quarks in 1+1 dimensions~\cite{1+1}.   A critical advantage of the $1+1$-dimensional system is that there exists a formalism that is known to become exact at large $N_c$ and can be formulated in a numerically tractable way using a lattice-based Hamiltonian formalism for arbitrary quark masses and densities~\cite{barak1,barak2}.  This enabled a direct comparison between the calculation based on the mean-field wave function with no hidden color states with the numerical approach of Ref.~\cite{barak1,barak2} computed with numerically large quark masses and numerically small densities.  As shown in Ref.~\cite{1+1}, the two approaches agreed up to expected numerical precision in the relevant regime, suggesting that mean-field approach for baryonic matter was valid for large $N_c$ and heavy quarks.  

The goal of this paper is to demonstrate the validity of the mean-field approximation for baryonic matter---which implicitly means no hidden color components---for the Witten limit of QCD with heavy quarks at large $N_c$.  We do this by taking the heavy quark limit at the outset, thus reducing the problem to one of pure quantum mechanics with color-Coulomb interactions with the proviso that the state is color-singlet.  In the remainder of this paper, we refer to the ``exact Hamiltonian'' as that color-Coulomb interaction and the ``exact wave function'' as the ground state of that Hamiltonian.  The question under consideration is whether the interaction energy of finite density baryonic matter for that Hamiltonian is well-approximated at large $N_c$ by a mean-field description.  We do this in the language of a generalized coherent state formalism.  This is natural because such a formalism has been used at a field theoretic level to establish formal properties of large $N_c$ QCD, demonstrating its classical or mean-field nature~\cite{yaffe}.  The generalized coherent states used here are, in fact, single Slater determinants~\cite{blaizot}; such states can be viewed as eigenstates of some effective mean-field Hamiltonian.  Thus, the question amounts to whether the exact energy is well-approximated at large $N_c$ by a single Slater determinant.

Two properties of the generalized coherent states are critical to the analysis.  The first is that they form an overcomplete basis \cite{blaizot}.  Thus, \textit{any} state---including the exact ground state---can be written as an integral over the parameters specifying the coherent state of a weight function times the coherent state.  There is an important corollary to this: all color-singlet states can be written as an integral over color-singlet coherent states.  Overcompleteness means that there is an infinite number of equivalent ways to do this. A second key property is that color-singlet generalized coherent states become exponentially narrow in parameter space as $N_c$ becomes large.  In other wods, the overlap between two different color-singlet coherent states is a number with norm less than unity raised to the power of $N_c$.  This implies that at large $N_c$ integrals over coherent states will be dominated by contributions which are nearly diagonal in the parameter space.

Exploiting these properties of the generalized coherent states, we consider the nature of the ground state for both the exact Hamiltonian of the many-body problem and the optimal mean-field Hamiltonian containing only one-body operators.  Our  approach is formally valid for arbitrary number of baryons in a fixed spatial region and, as such can describe finite density systems. We show that the many-body state that minimizes the exact Hamiltonian has an optimum width in the parameter space of the generalized coherent states. In other words, the state that minimizes the energy is narrowly peaked around a single coherent state.   As a result, the energy of the exact ground state is well-approximated by the energy of a single coherent state.  We further show that the exact ground state energy differs from the mean-field energy by an amount of relative order $1/N_c$.   

It is important to understand what this means for hidden color states.   Our generalized coherent states---i.e. single Slater determinant mean-field states---by construction contain no hidden color state components.  This does not mean that the true ground state does not contain these components.  Integration over the generalized coherent states induces hidden color state components since the coherent states being integrated over correspond to different mean-fields.  Moreover, the fact that the mean-field state, which gives a good approximation for the ground state energy and contains no hidden color components, is a single coherent state does not imply that hidden color states do not have contributions in the true ground state at large $N_{c}$.  All it implies is that the inclusion of hidden color states only alters the ground state energy by an amount of relative order $N_{c}^{-1}$.

\section{Baryon Hamiltonian}
We begin with the form of the nonrelativistic, many-body Hamiltonian that encodes two-body quark interactions in the heavy quark mass limit. First, we specify some region of space and boundary conditions for the wave functions. One then defines the Hilbert space to be the set of color-singlet states with fixed baryon number, $B$, satisfying these boundary conditions.  For the study of a single baryon, the spatial region is unrestricted and the boundary conditions are that the wave functions die off at infinity sufficiently rapidly to ensure square integrability. Infinite nuclear matter at fixed density can be studied by making a large box of fixed shape with a large number of baryons in it.  Formally, infinite nuclear matter is obtained by taking the thermodynamic limit in which the volume, $V$, goes to infinity while the baryon number, $B$, goes to infinity with baryon density, $\rho=V/B$, held fixed.  An alternative method for studying nuclear matter is to exploit the fact that in either the large $N_c$ or the heavy quark limits, infinite nuclear matter is expected to form a crystal with a fixed number of baryons per unit cell.  One could then postulate the crystal structure, and choose one (or more) unit cells with fixed size and baryon number (typically there will be $2 N_f$ degenerate copies of baryons in each cell \cite{3+1}), imposing appropriate boundary conditions.  The crystal structure chosen should be the one yielding the lowest energy density. 

For our purposes---the demonstration that mean-field theory in the large $N_c$ limit for color-singlet states with fixed baryon number accurately reproduces the energy up to corrections of relative order $1/N_c$---the detailed form of the Hamiltonian is irrelevant.  All that matters is that the Hamiltonian is color-singlet, contains only one-body operators (which we will take to consists of a mass operator and a kinetic energy operator) and two-body operators (which encode the interactions) with the one body operators having strength of order $N_{c}^{0}$ and the two-body operators with strength is of order $1/N_{c}$.

We write the Hamiltonian in terms of creation and annihilation operators for a set of single-quark operators:
\begin{equation}
\begin{split}
\{\hat{a}_\alpha^a,\hat{a}_\beta^b\}&=0\\
\{\hat{a}_\alpha^{\dagger a},\hat{a}_\beta^{\dagger b}\}&=0\\  
\{\hat{a}_\alpha^{ a},\hat{a}_\beta^{\dagger b}\}&=\delta_{\alpha\beta} \, \delta^{ab} \; .
 \label{sqo}\end{split}
\end{equation}
The Roman indices indicate color, while the Greek indices represent an arbitrarily chosen complete set of orthonormal single particle levels in space, spin and flavor.
The baryon Hamiltonian in second-quantized notation has the following form:
\begin{equation}\begin{split}
\label{Hamiltonian}
\hat{H}&= \hat{M}+ \hat{T}+\hat{V}\\
\hat{M}&=\sum_\alpha \sum_{a=1}^{N_c} M_Q \, \hat{a}_\alpha^{\dagger a} \hat{a}_\alpha^{ a}\\
\hat{T}&=\sum_{\alpha\beta}  \sum_{a=1}^{N_c} {t}_{\alpha\beta} \, \hat{a}^{\dagger a}_{\alpha}\hat{a}_{\beta}^a\\
\hat{V}&=\sum_{\alpha\beta\gamma\delta} \sum_{a=1}^{N_c}  \sum_{b=1}^{N_c} \frac{{v}_{\alpha\beta\gamma\delta}}{N_c} \,  \hat{a}^{\dagger a}_{\alpha}\hat{a}^{\dagger b}_{\beta}\hat{a}_{\gamma}^b \hat{a}_{\delta}^a
\end{split}\end{equation}
where $\hat{M}$ is the mass operator ,  $\hat{T}$ is the  kinetic energy operator and $\hat{V}$  is the potential energy operator.  $\hat{M}$ and  $\hat{T}$ are one-body operators while $\hat{V}$  is a two-body operators.   $M_{Q}$ is the quark mass. The coefficients ${t}_{\alpha\beta}$ and  ${v}_{\alpha\beta\delta\gamma}$ define the dynamics and are of order $N_{c}^{0}$.  Thus, for example, the spin-flavor independent color-Coulomb nature of the interaction is captured in the ${v}_{\alpha\beta\delta\gamma}$. However, as noted above, the precise form of the interaction is irrelevant to the question of whether mean-field theory is valid at large $N_c$.

\section{Generalized Coherent States}
Our primary concern in this paper is to understand the nature of the ground state for baryons and baryonic matter in large $N_{c}$ QCD and the relation between the mean-field, suggested by Witten~\cite{witten}, and the exact solution of the quantum mechanical problem.  Generalized coherent states provide a natural way to do this.  The mean-field approach can be cast as a variational problem; the mean-field ground state is the minimum energy solution within a particular class of trial functions. The variational space is the set of generalized coherent states of color-singlet states that carry the appropriate baryon charge and subject to the boundary conditions.  These generalized coherent states are, in fact, single Slater determinants.   Thus, the standard Hartree-Fock mean-field solution is the expectation value in one of these states.   In this section, we will review some important  properties of these generalized coherent states and develop these for the special case of color-singlet states.
 
The connection of these generalized coherent states to the usual coherent states is apparent when the generalized coherent states are written in an exponentiated form~\cite{thouless, blaizot, blaizotbook}  analogous to the standard Glauber states~\cite{klauder}. They differ in that instead of exponentiating a boson operator and acting it on the vacuum as in the Glauber state, one exponentiates a particle-hole operator in terms of the fermionic degrees of freedom and acts on a reference state.  As with the standard Glauber states, these generalized coherent states form an overcomplete basis.  Thus, the exact ground state can be written as an appropriate integral over these generalized coherent states \cite{blaizot, blaizotbook}.  Moreover, as we will prove below, the nature of this integral is such that if the ground state is a color-singlet, then all of the generalized coherent states over which the integral is done must also be color-singlet states.  Thus, the exact ground-state can be written as a weighted integral over color-singlet coherent states.
 
Before we construct the most general color-singlet coherent states of $N_{c}$ quarks and a baryon charge $B$, we define a reference state. All other generalized coherent states of the same baryon charge as the reference state will be non-orthogonal to it. The reference state with a baryon charge $B$ has the form:
\begin{equation}
\label{refstate}
|\textrm{ref},B\rangle=\prod_{h=1}^{B}\prod_{a=1}^{N_{c}}\hat{a}^{\dagger a}_{h}|\textrm{vac}\rangle\ .
\end{equation}
Here, the hole index $h=1\textrm{ to } B$, the color index $a=1 \textrm{ to } N_{c}$ and $|\textrm{vac}\rangle$ is the vacuum state, which has a baryon charge zero and a trivial structure since it is the vacuum of a non-relativistic theory. The operator $\hat{a}^{\dagger a}_{h}$ creates a quark with color $a$ in a ``hole'' state $h$. 
\begin{equation}
\hat{a}^{\dagger a}_{h}|\textrm{vac}\rangle=|q^{a}_{h}\rangle\ .
\end{equation}
All generalized coherent states will be created from $|\textrm{ref},B\rangle$ by removing occupied levels and creating particle-hole pairs relative to it.    The hole creation operators, $a_{h}^{\dagger a}$,  are a set of $B$ distinct $a_\alpha^{\dagger a}$ operators defined in Eq.~(\ref{sqo}).  All of the single levels that are not hole levels will be designated particle levels so that the anti-commutation relations become:
\begin{equation}
\label{anticomm}
\begin{split}
&\{\hat{a}^{\dagger a}_{p}, \hat{a}^{\dagger b}_{h} \}=\{\hat{a}_{p}^{a},\hat{a}^{\dagger b}_{h}\}=0\\
&\{\hat{a}_{p}^{a},\hat{a}^{\dagger b}_{p'}\}=\delta_{p p'}\delta^{a b} \textrm{ and } \\
& \{\hat{a}_{h}^{a},\hat{a}^{\dagger b}_{h'}\}=\delta_{h h'}\delta^{a b} \ .
\end{split}
\end{equation}
Note that the reference state is normalized and is a Slater determinant since all the creation operators in Eq.~(\ref{refstate}) anti-commute with each other. Furthermore, the  product over color ensures that the reference state is a color-singlet state.

All color-singlet generalized coherent states can be generated from the reference state in the following way:
\begin{equation}
\label{C}
|C,B\rangle=\mathcal{N}\exp\left(\sum_{ph}C_{ph}\sum_{a=1}^{N_{c}} \hat{a}^{\dagger a}_{p}\hat{a}^{a}_{h}\right ) |\textrm{ref},B\rangle \ .
\end{equation}
Here, the hole index, $h$, runs from $1\textrm{ to }B$ and the particle index, $p$,  goes from $B+1\textrm{ to }n_{p}+B$, where $n_{p}$ is the number of particle states. $C$ is a complex-valued, $n_{p}\textrm{ by }B$ matrix. $\mathcal{N}$ is the normalization factor, which can be easily determined. Here, we have assumed that the number of particle states is $n_{p}$, which for a finite volume system, is a countably infinite number. However, in the rest of the paper we will assume that $n_{p}$ is a finite but large number, in effect introducing an ultraviolet cutoff. As $n_{p}\rightarrow\infty$, one expects a smooth limit for all physical observables. Thus, approximating the system with a large but fintie value of $n_{p}$ is valid.

At first glance, it may not be obvious why the generalized  coherent states defined in Eq.~(\ref{C}) are the entire set of color-singlet, Slater determinants to the reference state. However, the state can be written in the following more transparent form using the anti-commutation relations of Eq.~(\ref{anticomm}):
\begin{equation}
\label{Cnew}
|C,B\rangle=\mathcal{N}\prod_{h=1}^{B}\prod_{a=1}^{N_{c}}\left( \hat{a}^{\dagger a}_{h}+\sum_{ph} C_{ph}\hat{a}^{\dagger a}_{p}\right )|\textrm{vac}\rangle \ .
\end{equation}
From the form stated above, it becomes quite clear that coherent states of Eq.~(\ref{C}) are indeed Slater determinants because the linear combination of creation operators of the form that appear in Eq.~(\ref{Cnew}): $\left(\hat{a}^{\dagger a}_{h}+\sum_{ph} C_{ph}\hat{a}^{\dagger a}_{p}\right  )$ with different color and hole indices all anti-commute with each other. Also, note that each linear combination of creation operators in Eq.~(\ref{C}) create many-body quark states of the following form:
\begin{equation}
|q_{h}^{a}\rangle+\sum_{ph}C_{ph}|q_{p}^{a}\rangle\ .
\end{equation}
Each of these states have a hole index, $h$, and a color index, $a$, and each combination of these indices only appear once. Therefore, the coherent states are all color-singlet states. Alternatively, the form of Eq.~(\ref{C}) should make clear that these states are color neutral since the reference state is color neutral (because the vacuum is trivial and colorless) and the exponential operator is by construction also color neutral.

Finally, it remains to be seen that the set of coherent states in Eq.~(\ref{C}) is  the set of all possible colorless Slater determinant states with fixed baryon number.  Strictly speaking this is not true, since from Eq.~(\ref{C}), it is clear that the state must contain a nonvanishing component of the reference state.  Thus, the state is non-orthogonal to the reference state.  However, by taking the limit in which the magnitude of $C$  goes to infinity, the normalization goes to zero and the state becomes orthogonal to the reference state. The most general color-singlet Slater determinant with baryon number $B$ obviously can be cast into the following form:
\begin{equation}
\label{generalstate}
|S,T,B\rangle=\mathcal{N'} \prod_{h=1}^{B}\prod_{a=1}^{N_{c}}\left ( \sum_{h'}S_{h'h}\hat{a}^{\dagger a}_{h'}+\sum_{p'}T_{p'h}\hat{a}^{\dagger a}_{p'} \right )| \textrm{vac}\rangle \ .
\end{equation}
Now we are considering the class of states which are non-orthogonal to the reference state $\langle\textrm{ref}|S,T,C\rangle\neq 0$. The reference state consists only of hole states, which implies that  $\det(S)\neq 0$.  This in turn means that $S$ is invertible and the state can be transformed easily to the coherent state of Eq.~(\ref{Cnew}) with $C=TS^{-1}$.

Now that we have established that the generalized coherent states of Eq.~(\ref{C}) are the complete set of all color-singlet Slater determinants with fixed $B$, we need to establish a few very important properties of these color-singlet generalized coherent states.   The first of these is the existence of an integral formula for the identity operator in our space:
\begin{equation}
\label{unity}
\begin{split}
\hat{1}_{\textrm{cs},B}& =\int d\mu(C,B)|\textrm{C},B\rangle\langle\textrm{C},B| \textrm{ with}\\
d\mu(C,B) &\equiv  J(C,B;N_{c},n_{p}) \prod_{p,h} dC_{ph}dC^{*}_{ph}\ .
\end{split}
\end{equation} 
$\hat{1}_{{\rm cs},B}$ is the unit operator for color-singlet states with baryon number, $B$;
$|\textrm{C},B\rangle$ is a normalized color-singlet generalized coherent state with baryon number $B$ of the form given in Eq.~(\ref{C}) and $d\mu(C,B)$ is the measure, which appropriately weights each coherent state and depends on the baryon number $B$, the number of colors, $N_{c}$ and the number of particle states, $n_{p}$, which is the ultraviolet cutoff and will be taken to infinity at the end of the problem. The explicit form of $J(\textrm{C},B;N_{c},n_{p})$ is given in Eq.~(\ref{jacobian}). 
 
The derivation of Eq.~(\ref{unity}) is straightforward.  One starts with a larger space---the space of states with a baryon number, $B$, but which is not limited to color-singlet states.  When discussing this larger space, a generic state will be denoted as $|\psi\rangle$ while a color-singlet state will be denoted  as $|\psi\rangle_{\rm cs}$.  When we return to a discussion of the color-singlet space only, we will drop the subscript ``cs".  The generalized coherent states with unit norm for this larger space are given by:
\begin{equation}
\label{D}
|D,B\rangle=\mathcal{N}_D\exp\left(\sum_{p a, h b}D_{p a, h b} \hat{a}^{\dagger a}_{p}\hat{a}^{b}_{h}\right ) |\textrm{ref},B\rangle ,
\end{equation}
where $\mathcal{N}_{D}$ is the normalization constant. Furthermore, the reference state, $|\textrm{ref}, B\rangle$, which has a baryon number, $B$, is assumed to be a color-singlet state. 

There is a well-known representation for the identity operator in this larger space~\cite{blaizot}.
\begin{equation}
\label{unity'}
\begin{split}
\hat{1}_{B}&=\int d\mu (D,B)|\textrm{D},B\rangle\langle\textrm{D},B|\ \textrm{with}\\
d\mu(D,B) &\equiv  J(D,B;N_{c},n_{p}) \prod_{pa,hb} dD_{pa,hb}dD_{pa,hb}^{*} \textrm{ and}\\
 J(D,B;N_{c},n_{p})&\equiv\frac{\mathcal{N}_{\text{ncs}}(B,n_{p},N_{c})}{\det(1+D^{\dagger}D)^{N_{c}(B+n_{p})}}\ .
\end{split}
\end{equation}
Here, it is important to note that the determinant is with respect to a combined particle-hole and color space. $\mathcal{N}_\textrm{ncs}$ is a numerical factor that depends on $B$, $n_{p}$ and $N_{c}$.

To proceed, we introduce the color-singlet projection operator $\hat{P}$ which projects onto color-singlet states. The operator has the following properties:
\begin{equation}
\begin{split}
&\textrm{i) }\hat{P}|\psi\rangle_\textrm{cs}=|\psi\rangle_\textrm{cs}\textrm{ for a color singlet state }|\psi\rangle_{\textrm{cs}}\\
&\textrm{ii) }\hat{P}  \hat{C}_{2}  \hat{P}=0\ ,
\end{split}
\label{progprops}
\end{equation}
where $\hat{C}_{2}$ is the quadratic Casimir operator for $SU(N_c)$.
Next we exploit condition i) of Eq.~(\ref{progprops}) to project out the color-singlet identity operator of Eq.~(\ref{unity}) using the more general representation of the identity operator in Eq.~(\ref{unity'}).
\begin{equation}
\hat{1}_{\textrm{cs},B}=\hat{P}\hat{1}_{B}\hat{P}=\int d\mu (D,B)\hat{P}|\textrm{D},B\rangle\langle\textrm{D},B|\hat{P}
\end{equation}
First, note that the action of the projection operator, $\hat{P}$, on the coherent state $|\textrm{D},B\rangle$ can be written in the following way:
\begin{equation}
\label{projectionofD}
\begin{split}
\hat{P}|\textrm{D},B\rangle&=\hat{P}\exp\left(\sum_{pa,hb}D_{pa,hb}\hat{a}^{\dagger a}_{p}\hat{a}^{b}_{h}\right)\hat{P}|\textrm{ref},B\rangle\\
&=\sum_{n=0}^{\infty}\frac{1}{n!}\mathcal{N}_{D}\hat{P}\left(\sum_{pa,hb}D_{pa,hb}\hat{a}^{\dagger a}_{p}\hat{a}^{b}_{h}\right)^{n}\hat{P}|\textrm{ref},B\rangle\ .
\end{split}
\end{equation}
In Eq.~(\ref{projectionofD}), the action of the color-singlet projection operator on the generalized coherent state, $|\textrm{D},B\rangle$, is to project out only the color-singlet component of the entire state. This is only possible if each term of the Taylor series expansion in Eq.~(\ref{projectionofD}) is separately a color-singlet. The first term with $n=0$ is a color-singlet since the reference state $|\textrm{ref},B\rangle$ is assumed to be a color-singlet state. The second term of the series, which is a one-particle one-hole term, can be a color-singlet only if the action of $\hat{P}$ is zero for all values of the matrix $D$, which are not of the form $D_{pa,hb}=\delta_{ab}C_{ph}$, where $C$ is a matrix only in particle-hole space and is independent of color.

Therefore, the identity operator of Eq.~(\ref{projectionofD}) can be written in the following way:
\begin{widetext}
\begin{equation}
\begin{split}
&\hat{1}_{\textrm{cs,B}}=\int \prod_{pa,hb} dD_{pa,hb}dD_{pa,hb}^{*} J(D,B;N_{c},n_{p})\hat{P}|\textrm{D},B\rangle\langle\textrm{D},B|\hat{P}\\
&=\int\prod_{ph}dC_{ph}dC^{*}_{ph}\int d\mu(\textrm{D},B)\prod_{p'h'}\delta\left(C_{p'h'}-\frac{1}{N_{c}}\sum_{ab}D_{p'a,h'b}\delta_{ab}\right)\delta\left(C^{*}_{p'h'}-\frac{1}{N_{c}}\sum_{ab}D^{*}_{p'a,h'b}\delta_{ab}\right)|\textrm{C},B\rangle\langle\textrm{C},B|\ ,
\end{split}
\end{equation}
\end{widetext}
where we have introduced integrals over $C_{ph}$ and $C^{*}_{ph}$ using appropriately normalized delta functions. This is the form of the identity operator in Eq.~(\ref{unity}) with the following identification for the function, $J(C,B;N_{c},n_{p})$:
\begin{widetext}
\begin{equation}
\label{jacobian}
J(C,B;N_{c},n_{p})=\int d\mu(\textrm{D},B)\prod_{p'h'}\delta\left(C_{p'h'}-\frac{1}{N_{c}}\sum_{ab}D_{p'a,h'b}\delta_{ab}\right)\delta\left(C^{*}_{p'h'}-\frac{1}{N_{c}}\sum_{ab}D^{*}_{p'a,h'b}\delta_{ab}\right)\ .
\end{equation}
\end{widetext}

Another important property of these set of generalized coherent states is the overlap between two distinct states. This can be determined quite easily using the standard determinant relation for the overlap of two fermionic states \cite{blaizotbook}:
\begin{equation}
\label{overlap}
\langle\textrm{C},B|\textrm{C}'B'\rangle=\delta_{BB'}\frac{\det(1+C^{\dagger}C')^{B N_{c}}}{\det(1+C^{\dagger} C)^{\frac{B N_{c}}{2}}\det(1+C'^{\dagger} C')^{\frac{B N_{c}}{2}}}\ .
\end{equation}
The determinant in the above equation is taken with respect to the particle-hole space only. The overlap of two coherent states with different baryon numbers is zero; the overlap of two identical coherent states is fixed by normalization to be one. It is useful to rewrite the overlap in terms of $\overline{C}=\frac{1}{2} (C+C')$ and $\Delta C=C-C'$.
\begin{widetext}
\begin{equation}\label{overlap'}\begin{split}
\langle\textrm{C},B|\textrm{C}'B'\rangle&=\langle \overline{C}+\Delta C/2,B |\overline{C}-\Delta C/2,B \rangle =\delta_{BB'} f\left (\overline{C},\ \Delta C\right )^{N_c} \\
f(\overline{C},\Delta C )&=  \frac{\det\left ( 1+\overline{C}^{\dagger}\overline{C}-\frac{\overline{C}^{\dagger}\Delta C}{2} + \frac{\Delta C^\dagger \overline{C}}{2} - \frac{\Delta C^\dagger \Delta C}{4} \right )}{\det(1+\overline{C}^{\dagger}\overline{C}+ \frac{\overline{C}^\dagger  \Delta C}{2}+\frac{\Delta C^\dagger \overline{C}}{2}+ \frac{\Delta C^\dagger \Delta C}{4} )^{\frac{1}{2}}   \det(1+\overline{C}^{\dagger}\overline{C}- \frac{\overline{C}^\dagger  \Delta C}{2}-\frac{\Delta C^\dagger \overline{C}}{2}+\frac{\Delta C^\dagger \Delta C}{4} )^{\frac{1}{2}}}\ .
\end{split}
\end{equation}
\end{widetext}
It is easy to see that $f(\overline{C},0)=1$ for all $\overline{C}$ and that $|f(\overline{C},\Delta C)|<1$ for all $\Delta C \neq 0$.  Thus, it is is clear that the overlap falls off exponentially with $N_c$ as $\Delta C$ deviates from zero.

The generalized coherent states form an overcomplete basis and therefore can be used to express any state in the space of color-singlet states. A state with baryon number $B$ can therefore be expressed as an integral involving color-singlet coherent states with the same baryon number:
\begin{equation}
|\psi\rangle=\int d\mu(C, B) w_{\psi}(C) |C,B\rangle\ .
\label{arb}\end{equation}
Here $w_{\psi}$ is a weight function but it is not unique.  It is easy to see from the form of the identity operators in Eq.~(\ref{unity}) and Eq.~(\ref{overlap}) that if $w_\psi^{(j)}(C)$ is a weight function for state $|\psi\rangle$, then so is
\begin{align}
\label{altform}
&w_\psi^{(j+1)}(C)=\\
&\int d\mu(C',B) \frac{w_{\psi}^{(j)}(C') \; \det(1+C'^{\dagger}C)^{B N_{c}}}{\det(1+C^{\dagger} C)^{\frac{B N_{c}}{2}}\det(1+C'^{\dagger} C')^{\frac{B N_{c}}{2}}} \ .
\nonumber \end{align}
Linear combinations of the form $\sum_{i} \alpha_{i}   w_{\psi}^{(i)}$ with $ \sum \alpha_{i}=1$ also yield valid weight functions.  The non-uniqueness of  $w_\psi$ reflects the overcompleteness of the generalized coherent states basis.  One can, however, define a primary weight function $w_\psi^{(0)}(C)$ as follows:
\begin{equation}
w_\psi^{(0)}(C)\equiv \langle C,B|\psi \rangle \ .
\end{equation}
It is straightforward to see from the form of the identity operator in Eq.~(\ref{unity}) that this is indeed a weight function for $|\psi\rangle$.

\section{The validity of mean-field theory}
Mean-field theory amounts to replacing the exact ground state with a single Slater determinant, $|C_{\textrm{mf}}\rangle$, which is the variational minimum in the space of color-singlet states with a finite baryon charge.  This state is not the solution of the exact Hamiltonian of Eq.~(\ref{Hamiltonian}), since the potential energy term contains two-body operators.   The goal of this paper is to provide a rigorous justification for the mean-field theory as providing an accurate calculation of the energy of the heavy quark Hamiltonian to leading order in the $1/N_c$ expansion.  In other words, we will show that using the mean-field Hamiltonian gives rise to corrections in energy that are of relative order $N_{c}^{-1}$:
\begin{equation}
\label{diff1}
\frac{\langle \psi_{\textrm{g.s.}}|\hat{H}|\psi_{\textrm{g.s.}}\rangle-\langle C_{\textrm{mf}}|\hat{H}|C_{\textrm{mf}}\rangle}{\langle \psi_{\textrm{g.s.}}|\hat{H}|\psi_{\textrm{g.s.}}\rangle}\sim\mathcal{O}(N_{c}^{-1})\ .
\end{equation}
We have assumed here that the states $|\psi_{\textrm{g.s.}}\rangle$ and $|C_{\textrm{mf}}\rangle$ are both appropriately normalized.

To proceed, we first note that the true ground state of the Hamiltonian in Eq.~(\ref{Hamiltonian}), which we denote $|\psi_{\textrm{g.s.}}\rangle$, can be written in terms of a weight function as in Eq.~(\ref{arb}) as follows:
\begin{widetext}
\begin{equation} 
\langle \psi_{\textrm{g.s.}}|\hat{H}|\psi_{\textrm{g.s.}}\rangle = \int \prod_{p,h,p',h'} dC_{ph} dC_{ph}^{*}dC'_{p'h'}dC'^{*}_{p'h'} \,  \left ( J(C,B) w_{\psi_{\textrm{g.s.}}} (C) \right )^* \left( J(C',B)   w_{\psi_{\textrm{g.s.}} }\!(C') \right ) \, { \langle C| \Hat{H} |C' \rangle}\ ,
\end{equation}
where we have used the norm in Eq.~(\ref{unity}). It is useful to change variables in this integral in terms of $\overline{C}$ and $\Delta C$:
\begin{equation}
\begin{split}
\label{gform}
\langle \psi_{\textrm{g.s.}}|\hat{H}|\psi_{\textrm{g.s.}}\rangle& = N_c \int \prod_{p,h,p',h'} d\overline{C}_{ph}d\overline{C}^{*}_{ph}d\Delta C_{p'h'}d\Delta C^{*}_{p'h'} \ G_{{\psi_{\textrm{g.s.}} }}\!(\overline{C},\Delta C) h(\overline{C},\Delta C),\textrm{where} \\
G_{{\psi_{\textrm{g.s.}} }}\!(\overline{C},\Delta C)& =\left ( J(\overline{C} +\Delta C/2,B) w_{\psi_{\textrm{g.s.}}} (\overline{C} +\Delta C/2) \right )^* \left( J(\overline{C} -\Delta C/2,B)   w_{\psi_{\textrm{g.s.}} }\!(\overline{C} -\Delta C/2) \right ) f(\overline{C},\Delta C )^{N_c},\ \textrm{and}\\
h(\overline{C},\Delta C) &\equiv \frac{ \langle \overline{C},+\Delta C/2|\hat{H}|\overline{C}-\Delta C/2 \rangle}{N_c \langle \overline{C}+\Delta C/2|\overline{C}-\Delta C/2\rangle} \; 
\end{split}
\end{equation}
\end{widetext}
with $f(\overline{C},\Delta C )$ defined in Eq.~(\ref{overlap'}).   Note that by construction $ G_{{\psi_{\textrm{g.s.}} }}\!$ is normalized:
\begin{equation}
 \int \prod_{p,h,p',h'} d\overline{C}_{ph}d\overline{C}^{*}_{ph}d\Delta C_{p'h'}d\Delta C^{*}_{p'h'} \, G_{{\psi_{\textrm{g.s.}} }}\!(\overline{C},\Delta C)=1\ .
\end{equation}

The function, $h(\overline{C},\Delta C)$, completely characterizes the Hamiltonian. It is of the form:
\begin{equation}
\label{hhh}
\begin{split}
h(\overline{C},\Delta C)&=h_{0}(\overline{C},\Delta C) + \frac{1}{N_c}h_{1}(\overline{C},\Delta C)\ ,
\end{split}
\end{equation}
where both $h_{0}$ and $h_{1}$ are $\mathcal{O}(N_{c}^{0})$ functions. The explicit forms of both $h_{0}$ and $h_{1}$ are presented in Appendix~(\ref{h}).

The important thing to note for the remainder of the analysis is that $h_0$ and $h_1$ are both independent of $N_c$.  This means that these two functions remain smooth functions of $\Delta C$ even in the limit $N_{c} \rightarrow \infty$. 

In contrast, $G_{{\psi_{\textrm{g.s.}} }}\!(\overline{C},\Delta C)$ contains a factor of  $f(\overline{C},\Delta C )^{N_c}$ and thus one expects it to be exponentially narrow as a function of $\Delta C$ at large $N_c$. To emphasize this point, it is useful to define the following function
\begin{equation}
\label{defineg}
g_{{\psi_{\textrm{g.s.}}}}\!(\overline{C}) \equiv \int \prod_{p,h}  d\Delta C_{ph}d\Delta C_{ph}^{*} \, G_{{\psi_{\textrm{g.s.}} }}\!(\overline{C},\Delta C) \ .
\end{equation}
Then the function, $G_{{\psi_{\textrm{g.s.}} }}\!(\overline{C},\Delta C)$, can be rewritten in the following form:
\begin{equation}
\begin{split}
G_{{\psi_{\textrm{g.s.}} }}\!(\overline{C},\Delta C) &= g_{{\psi_{\textrm{g.s.}} }}\!(\overline{C}) \prod_{ph} \delta(\Delta C_{ph})\delta(\Delta C_{ph}^{*}) \, \\
&+ \, \Delta G_{{\psi_{\textrm{g.s.}} }}\!(\overline{C},\Delta C)\ .
\label{defineG}
\end{split}
\end{equation}
Note that $g_{{\psi_{\textrm{g.s.}} }}\!(\overline{C})$ is defined so that it has the properties of a probability: it is non-negative for all $\overline{C}$ and integrates to unity. $ \Delta G_{{\psi_{\textrm{g.s.}} }} \!(\overline{C},\Delta C) $, which appears in Eq.~(\ref{defineG}) has the property that it integrates to zero and is exponentially peaked in $\Delta C$.   

Let us rewrite Eq.~(\ref{gform}) using the form in  Eq.~(\ref{defineg}):
\begin{widetext}
\begin{equation}
\begin{split}
\label{gform2}
&\langle \psi_{\textrm{g.s.}}|\hat{H}|\psi_{\textrm{g.s.}}\rangle = E^{(1)}_{\psi_{\textrm{g.s.}}}+E^{(2)}_{\psi_{\textrm{g.s.}}}\textrm{ with}\\
&\frac{E^{(1)}_{\psi_{\textrm{g.s.}}} }{N_c}=  \int \prod_{p,h} d\overline{C}_{ph}d\overline{C}_{ph}^{*} \, g_{{\psi_{\textrm{g.s.}} }}\!(\overline{C}) h(\overline{C},0)\\
&\frac{E^{(2)}_{\psi_{\textrm{g.s.}}} } {N_c}= \int \!\! \prod_{p,h,p',h'} d\overline{C}_{ph}d\overline{C}_{ph}^{*} d\Delta C_{p'h'}d\Delta C_{p'h'}^{*}\Delta G_{{\psi_{\textrm{g.s.}} }}\!(\overline{C},\Delta C) h(\overline{C},\Delta C)\ .
\end{split}
\end{equation}
\end{widetext}

The values of $E^{(1)}_{\psi_{\textrm{g.s.}}}$ and $E^{(2)}_{\psi_{\textrm{g.s.}}}$ depend implicitly on the weight function, $w_{\psi_{\textrm{g.s.}} }$, since it goes into the definition of $G_{\psi_{\textrm{g.s.}} }$.  As shown in Eq.~(\ref{altform}), $w_{\psi_{\textrm{g.s.}} }$ is not uniquely defined.  Thus, neither are $E^{(1)}_{\psi_{\textrm{g.s.}}}$ and $E^{(2)}_{\psi_{\textrm{g.s.}}}$.  However, their sum is fixed; changes in $E^{(1)}_{\psi_{\textrm{g.s.}}}$ induced by an alternative form for $w_{\psi_{\textrm{g.s.}} }$ are precisely compensated for by changes in $E^{(2)}_{\psi_{\textrm{g.s.}}}$.

Note that $h(\overline{C},0)$ is, by definition, the expectation value (per color) of the Hamiltonian with respect to a generalized coherent state. By definition, the minimum value of  $N_{c}\,h(\overline{C},0)$ is $\langle C_{\textrm{mf}}|\hat{H} |C_{\textrm{mf}}\rangle$.  Since $ \, g_{{\psi_{\textrm{g.s.}} }}\!(\overline{C})$ is not negative and integrates to unity, it follows that 
\begin{equation}
\label{E1}
E^{(1)}_{\psi_{\textrm{g.s.}}} \ge \langle C_{\textrm{mf}}|\hat{H} |C_{\textrm{mf}}\rangle \; .
\end{equation}
Since we expect $G_{{\psi_{\textrm{g.s.}} }}\!(\overline{C},\Delta C)$  to be exponentially narrow  as a function of $\Delta C$ at large $N_c$, it can be well-approximated by a Gaussian of characteristic width of order $N_c^{1/2}$ inside of integrals with respect to $\Delta C$ and $\Delta C^{*}$.  Using this, one finds that integrals with respect to $\Delta C$ and $\Delta C$ of $\Delta G$ times a smoothly varying function (such as $h(\overline{C},\Delta C)$) will be of order $1/N_c$ implying that $E^{(2)}_{\psi_{\textrm{g.s.}}}$ is of order $N_{c}^0$.  Using this fact along with Eqs.~(\ref{gform2}) and (\ref{E1}) implies
\begin{equation}
\label{final}
\langle \psi_{\textrm{g.s.}}|\hat{H}|\psi_{\textrm{g.s.}}\rangle \ge  \langle C_{\textrm{mf}}|\hat{H} |C_{\textrm{mf}}\rangle + {\cal O}(N_c^{0})
\end{equation}
On the other hand, $\langle \psi_{\textrm{g.s.}}|\hat{H}|\psi_{\textrm{g.s.}}\rangle \le \langle C_{\textrm{mf}}|\hat{H} |C_{\textrm{mf}}\rangle$, by definition since $|\psi_{\textrm{g.s.}}\rangle$ is the ground state.  This is possible only if the $\mathcal{O}(N_c^{0})$ is negative and therefore we obtain Eq.~(\ref{diff1}).  This is the principle result of this paper.  We have demonstrated that the mean-field expression for the energy is valid up to corrections of relative order $1/N_c$.

There remains one possible subtlety.  In this derivation, we used the fact that $G_{{\psi_{\textrm{g.s.}} }}\!(\overline{C},\Delta C)$  as a function of $\Delta C$ is well-approximated by a Gaussian with a characteristic width of order $N_c^{-1/2}$.  This followed from the factor of $ f(\overline{C},\Delta C )^{N_c}$ contained in $G_{{\psi_{\textrm{g.s.}} }}\!(\overline{C},\Delta C)$.  However,	 $G_{{\psi_{\textrm{g.s.}} }}\!(\overline{C},\Delta C)$ contains other factors.  It is easy to see that these cannot make $G_{{\psi_{\textrm{g.s.}} }}\!(\overline{C},\Delta C)$ characteristically wider.  However, one might worry that an exceptionally narrow weight function $w_{\psi_{\textrm{g.s.}}}$ could lead to $G_{{\psi_{\textrm{g.s.}} }}\!(\overline{C},\Delta C)$, which is characteristically narrower than order $N_c^{-1/2}$ in $\Delta C$ in the regime of $\overline{C}$ which dominates; this could change the result.  For example, if it were to have turned out that $w_{\psi_{\textrm{g.s.}}}$ had a characteristic width of $N_c^{-1}$, the correction term in Eq.~(\ref{final}) would be $\mathcal{O}(N_c^{-1})$.  Since the correction term is negative this would be less bound than is given in Eq.~(\ref{diff1}) and can thus be ruled out for the ground state.

In summary, we have used the properties of generalized coherent states to show that mean-field theory accurately gives the energy for this system at large $N_c$ up to corrections of relative order $1/N_c$.  The result is precisely what was expected for finite density baryon matter for QCD with heavy quarks and many colors, but it is important to show that the expectations are correct.  In particular, it is encouraging to see that the result holds despite the possibility of hidden color components in the wave function.

\appendix
\begin{widetext}
\section{$h_{0}$ and $h_{1}$}

In this appendix, we present the explicit forms of $h_{0}$ and $h_{1}$. They are defined in Eqs.~(\ref{gform}) and (\ref{hhh}).
\begin{equation}
\label{h}
\begin{split}
h_0(\overline{C},\Delta C)
&+\sum_{\alpha\beta hh'}(M_{Q}\delta_{\alpha\beta}+t_{\alpha\beta})\left (1+\overline{C}-\frac{\Delta C}{2}\right )_{\alpha h}\left ( 1+\overline{C}^{\dagger}\overline{C}-\frac{\overline{C}^{\dagger}\Delta C}{2}+\frac{\Delta C^{\dagger}{\overline{C}}}{2} -\frac{\Delta C^{\dagger}\Delta C}{4}\right )^{-1}_{hh'}\left ( 1+\overline{C}^{\dagger}+\frac{\Delta C^{\dagger}}{2} \right )_{h'\beta}\\
&+\sum_{\alpha\beta\gamma\delta}v_{\alpha\beta\gamma\delta}\left(1- \left (1+\overline{C}-\frac{\Delta C}{2}\right )\left ( 1+\overline{C}^{\dagger}\overline{C}-\frac{\overline{C}^{\dagger}\Delta C}{2}+\frac{\Delta C^{\dagger}{\overline{C}}}{2} -\frac{\Delta C^{\dagger}\Delta C}{4}\right )^{-1}\left ( 1+\overline{C}^{\dagger}+\frac{\Delta C^{\dagger}}{2} \right )\right)_{\alpha\delta}\\
& \left(1- \left (1+\overline{C}-\frac{\Delta C}{2}\right )\left ( 1+\overline{C}^{\dagger}\overline{C}-\frac{\overline{C}^{\dagger}\Delta C}{2}+\frac{\Delta C^{\dagger}{\overline{C}}}{2} -\frac{\Delta C^{\dagger}\Delta C}{4}\right )^{-1}\left ( 1+\overline{C}^{\dagger}+\frac{\Delta C^{\dagger}}{2} \right )\right)_{\beta\gamma}\\
&+\sum_{\alpha\beta\gamma\delta}v_{\alpha\beta\gamma\delta}\left (\left (1+\overline{C}-\frac{\Delta C}{2}\right )\left ( 1+\overline{C}^{\dagger}\overline{C}-\frac{\overline{C}^{\dagger}\Delta C}{2}+\frac{\Delta C^{\dagger}{\overline{C}}}{2} -\frac{\Delta C^{\dagger}\Delta C}{4}\right )^{-1}\left ( 1+\overline{C}^{\dagger}+\frac{\Delta C^{\dagger}}{2} \right )\right )_{\alpha\delta}1_{\beta\gamma}\\
&+\sum_{\alpha\beta\gamma\delta}v_{\alpha\beta\gamma\delta}1_{\alpha\delta}\left (\left (1+\overline{C}-\frac{\Delta C}{2}\right )\left ( 1+\overline{C}^{\dagger}\overline{C}-\frac{\overline{C}^{\dagger}\Delta C}{2}+\frac{\Delta C^{\dagger}{\overline{C}}}{2} -\frac{\Delta C^{\dagger}\Delta C}{4}\right )^{-1}\left ( 1+\overline{C}^{\dagger}+\frac{\Delta C^{\dagger}}{2} \right )\right )_{\beta\gamma}\\
&-1_{\alpha\delta}1_{\beta\gamma}
\end{split}
\end{equation}
\end{widetext}
\begin{widetext}
\begin{equation}
\begin{split}
h_1(\overline{C},\Delta C)=&-\sum_{\alpha\beta\gamma\delta}v_{\alpha\beta\gamma\delta}\left(1- \left (1+\overline{C}-\frac{\Delta C}{2}\right )\left ( 1+\overline{C}^{\dagger}\overline{C}-\frac{\overline{C}^{\dagger}\Delta C}{2}+\frac{\Delta C^{\dagger}{\overline{C}}}{2} -\frac{\Delta C^{\dagger}\Delta C}{4}\right )^{-1}\left ( 1+\overline{C}^{\dagger}+\frac{\Delta C^{\dagger}}{2} \right )\right)_{\alpha\gamma}\\
& \left(1- \left (1+\overline{C}-\frac{\Delta C}{2}\right )\left ( 1+\overline{C}^{\dagger}\overline{C}-\frac{\overline{C}^{\dagger}\Delta C}{2}+\frac{\Delta C^{\dagger}{\overline{C}}}{2} -\frac{\Delta C^{\dagger}\Delta C}{4}\right )^{-1}\left ( 1+\overline{C}^{\dagger}+\frac{\Delta C^{\dagger}}{2} \right )\right)_{\beta\delta}\\
&-\sum_{\alpha\beta\gamma\delta}v_{\alpha\beta\gamma\delta}\left (\left (1+\overline{C}-\frac{\Delta C}{2}\right )\left ( 1+\overline{C}^{\dagger}\overline{C}-\frac{\overline{C}^{\dagger}\Delta C}{2}+\frac{\Delta C^{\dagger}{\overline{C}}}{2} -\frac{\Delta C^{\dagger}\Delta C}{4}\right )^{-1}\left ( 1+\overline{C}^{\dagger}+\frac{\Delta C^{\dagger}}{2} \right )\right )_{\alpha\gamma}1_{\beta\delta}\\
&-\sum_{\alpha\beta\gamma\delta}v_{\alpha\beta\gamma\delta}1_{\alpha\gamma}\left (\left (1+\overline{C}-\frac{\Delta C}{2}\right )\left ( 1+\overline{C}^{\dagger}\overline{C}-\frac{\overline{C}^{\dagger}\Delta C}{2}+\frac{\Delta C^{\dagger}{\overline{C}}}{2} -\frac{\Delta C^{\dagger}\Delta C}{4}\right )^{-1}\left ( 1+\overline{C}^{\dagger}+\frac{\Delta C^{\dagger}}{2} \right )\right )_{\beta\delta}\\
&+1_{\alpha\gamma}1_{\beta\delta}\ .
\end{split}
\end{equation}
Here $\alpha$, $\beta$, $\gamma$ and $\delta$ represent either particle or hole indices. Furthermore, note that $\overline{C}_{\alpha h}$ and $\Delta C_{\alpha h}$ are zero if $\alpha$ is a hole index. 
\end{widetext}

\begin{acknowledgements}
P.A. and T.D.C. acknowledge the support of the U.S. Department of Energy through grant number DEFG02-93ER-40762.
\end{acknowledgements}


\begin{thebibliography}{20}
\bibitem{witten} E. Witten, Nucl. Phys. B \textbf{160}, 57 (1979).
\bibitem{signproblem1} I. M. Barbour, S. E. Morrison, E. G. Klepfish, J. B. Kogut, M.-P. Lombardo and UKQCD Collaboration, Nucl. Phys. B \textbf{60A} 220 (1998).
\bibitem{signproblem2} I.M. Barbour, S.E. Morrison, E.G. Klepsh, J.B. Kogut and M.P. Lombardo, Nucl. Phys. Proc. Suppl. \textbf{60A}, 220 (1998).
\bibitem{AS1} E. Corrigan and P. Ramond, Phys. Lett. B \textbf{87}, 73 (1993).
\bibitem{AS2} Thomas D. Cohen, Daniel L. Shafer and Richard F. Lebed, Phys. Rev. D \textbf{81}, 3 (2010).
\bibitem{3+1} T.D. Cohen, N. Kumar and K. K. Ndousse, Phys. Rev. C \textbf{84}, 015204 (2011).
\bibitem{hidden1} S. J. Brodsky, C.-R. Ji and G.P. Lepage, Phys. Rev. Lett. \textbf{51}, 83 (1983).
\bibitem{hidden2} D.M. Brink and Fl. Stancu, Phys. Rev. D \textbf{49}, 4665 (1994).
\bibitem{hidden3} M. Harvey, Nucl. Phys. A, \textbf{352}, 301 (1981). 
\bibitem{1+1} P. Adhikari, A. T.D. Cohen, A. Jamgochian, N. Kumar, Phys. Rev. C $\textbf{87}$ 035205 (2012); arXiv:nucl-th/1212.2167 (2012).
\bibitem{barak1} B. Bringoltz, Phys. Rev. D \textbf{79}, 105021 (2009). 
\bibitem{barak2} B. Bringoltz, Phys. Rev. D \textbf{79}, 125006 (2009).
\bibitem{yaffe} L. G. Yaffe, Rev. Mod. Phys. 54, 407-435 (1982).
\bibitem{blaizot} J.P. Blaizot and H. Orland, Phys. Rev. C \textbf{24} (1981).
\bibitem{thouless} D.J. Thouless, Nucl. Phys. \textbf{21} (1960).
\bibitem{blaizotbook} Blaizot, J-P, and Ripka, G. \textit{Quantum Theory of Finite Systems}, (MIT Press, Cambridge, London, 1986).
\bibitem{klauder} John R. Klauder and E.C.G. Sudarshan  \textit{Quantum Optics}, (Benjamin, New York, Amsterdam, 1968).



\end{thebibliography}
\end{document}